# CONSTRUCTION AND CHARATERIZATION OF SYMMETRICAL STATES FOR MULTIQUBIT SYSTEMS

P. J. Lin-Chung
Naval Research Laboratory
Washington, D. C. 20375

## Abstract

A general method in constructing a complete set of wave functions for multipartite identical qubits is presented based on the irreducible representations of the permutation group and the *nth* rank tensors. Particular examples for $n = 2$, 3, and 4 are derived and the entanglement behavior for each state is examined from several criteria. It is found that the states so constructed are all bound entangled states. For the case of even $n$, all the states are found to have maximum "$n$-tangle". The symmetry in spin space is found to increase the $n$-tangle in general. The "$n$-tangle" for $n = 4$ is found not always representing 4-way entanglement. It measures the degree of spin-space symmetry instead. A useful relationship in the classification between systems containing different number of qubits is given in terms of the Young's Tableaux based on our analysis.

I. Introduction

Quantum entanglement has puzzled physicists throughout decades of modern quantum mechanics. It also becomes a significant resource in the modern quantum computation, quantum teleportation, and quantum encryptus [1]. Multipartite entanglement is especially important in the study of decoherence, dense coding, and error detection in quantum computers [2] and in the study of secret sharing, quantum cloning, and controlled coding in telecommunication and telecomputing [3-7]. In addition, the multipartite entangled states have many nonclassical correlation physics waiting to be explored [2]. The simpler bipartite entanglement has been extensively examined previously [8, 9]. There are still many open questions remained to be answered after many years of progress to understand its nature. Among the questions are the measurements of the entanglement, the definitions of degree of entanglement, and the maximal entanglement. The construction and classification of entangled states for systems containing more than two qubits have not been studied in the same level of sophistication as the bipartite states. Tri-partite [10-14] and 4-partite [12, 15-16 ] systems have been received much attention recently. All the attention has been concentrated at general mixed states, which do not possess the symmetry of identical qubits presumably due to the technological difficulty in producing them and due to the interaction with the environment.

Experimentally the preparation of desired entangled multipartite states involves the design of quantum circuit with sequence of Hadamard gate and pairwise acting C-NOT gates [17]. Cavity quantum electrodynamics was also proposed using current technologies to prepare the $n$-qubit W states [18]. With the advance of technology it becomes feasible to prepare multipartite systems of any given design. A systematic theoretical study of the multipartite entanglements in a complete set of states for identical qubit system therefore becomes more important for deeper understanding of the dynamics in many-body quantum systems.

The present work provides a general method to construct the wave functions of multipartite identical qubits based on the symmetry of the permutation groups, and to classify these states according to their special irreducible representations in the symmetry groups. The states so obtained have the desired symmetry in spin space as well as in the phase-space, and are shown to be all pure and entangled. The separability and the degree of entanglement associated with each irreducible representation are examined from several criteria [14, 19, 20]. A useful relationship of the entanglements between systems containing different number of qubits is given in terms of the Young's tableau [21] based on our analysis.

The organization of this paper is as follows. Sec. II will describe the method of constructing the irreducible representations of the permutation group and the n-th rank tensors. Sec. III gives the examples of bipartite Bell States, tripartite states, and 4-qubit states. Summary and discussion will be given in Sec. IV.

II. Method

(1) Construction of symmetrical basis in tensor space.

For an identical *n*-particle system in *m*-dimensional space $V_m$, the direct product space, $V_m^n \equiv V_m \times V_m \times V_m \times V_m \times \cdots$, forms a tensor space with basis

$$|i\rangle_n \equiv |i_1, i_2, \cdots, i_n\rangle = |i_1\rangle|i_2\rangle \cdots |i_n\rangle , \quad \text{Where } |i_l\rangle \text{ is } m\text{-dimensional.} \tag{1}$$

We aim to construct symmetric basis, which remain unchanged by permutations of individual particles. This can be done through the consideration of the permutation symmetry group $S_n$. Under certain circumstances the basis states are also required to have special symmetry in the $V_m$ space. An example is that for $m = 2$ we want to obtain the symmetric and anti-symmetric state basis such as the GHZ entangled states [22]. The present approach is to construct the tensor basis with the symmetry of $S_n$ first. Then using the new basis sets to construct the final basis that contain the required property in the $V_m$ space.

The permutation group $S_n$ describes the symmetry of the $n$ identical particles. There is a special elegant way to determine the classes of the $S_n$ group using Young's Diagrams, which are the graphical representations of the different partitions associated with $S_n$ [21]. Each diagram consists of $n$ squares arranged in rows of different width. The number of distinct Young's Diagrams for a given $n$ is equal to the number of classes in $S_n$ and it in turn is equivalent to the number of inequivalent irreducible representations of $S_n$. By filling the squares of a Young's diagram with letters $a$, $b$, $c$, …. , in order from left to right and from the top row to the bottom row one obtains a normal Young's Tableau. In analogy to the construction of basis functions for irreducible representations in the space groups in solid state physics using the basis function generation machine [23], the tensors of the symmetry class λ consisting $e_\lambda |i\rangle_n$, where $|i\rangle_n \in V_m^n$, can be generated from each Tableau λ by first identifying the irreducible symmetrizer, $e_\lambda$, which is the primitive idempotent associated with the symmetry class. Using the basis function generating machine, the simple application of the operation $e_\lambda$ on any one of the basis $|i\rangle_n$ will give either zero or one of the basis functions associated with the λth irreducible representation of the permutation group $S_n$ in $V_m^n$ space. The normalization factor is not given in this generation procedure, but is easily added afterward.

The basis functions so obtained may be designated as "permutation harmonics" in analogy to the "cubic harmonics" generated from the crystal space groups. They are related to the components of the total angular momentum of the system. The Young's Tableaux also serve to relate the evolution of irreducible representations for different number of $n$ in a transparent way as we shall see in our examples in Sec. III.

Because the irreducible representation dictates the specific outcome of symmetry transformations its basis vectors would undergo, and because the basis functions of different irreducible representations are orthogonal, it is a natural way to classify the wave functions in $V_m^n$ space according to the representation they belong.

In additional to the permutation symmetry in *n* space, the transformation within the *m* dimensional space of each particle must also be considered. We shall only consider the *m* = 2 case in the following discussion for qubits.

If $\{g\}$ represents a set of linear transformations on $V_m$, and if, with respect to the law of multiplication, $\{g\}$ forms a group, $G_m$, in $V_m$, we have

$$g|a\rangle = \sum_{b=1}^{m} |b\rangle g_a^b. \tag{2}$$

This group $G_m$ induces $n \times m$ dimensional representation of transformations in $V_m^n$ space, such as,

$$D(g)_i^j \equiv g_{i_1}^{j_1} g_{i_2}^{j_2} g_{i_3}^{j_3} \cdots g_{i_n}^{j_n} \tag{3}$$

In $V_2$ space the Pauli spin operator $\sigma_x$ is known to be a local spin-flip (or bit flip) operator and $\sigma_z$ is the phase flip operator. If we use $\sigma_{xj}$ to represent the spin flip operator in the *j* subspace in $V_2^n$ (*i.e.* the *j*-th qubit), and the *j*th qubit has the wave function $a_0|0\rangle + a_1|1\rangle$, here $|0\rangle, |1\rangle,$ label the spin up and spin down states, then

$$\sigma_{xj} \begin{vmatrix} a_0 \\ a_1 \end{vmatrix} = \begin{vmatrix} a_1 \\ a_0 \end{vmatrix} \tag{4a}$$

$$\sigma_{zj} \begin{vmatrix} a_0 \\ a_1 \end{vmatrix} = \begin{vmatrix} a_0 \\ -a_1 \end{vmatrix} \tag{4b}$$

After the distinct basis vectors associated with each irreducible representation of $S_n$ in $V_2^n$ space are found, the entangled states which are symmetric or anti-symmetric in spins are obtained by applying the operator $T_\pm^{(n)} \equiv \frac{1}{\sqrt{2}} (1 \pm \sigma_{xa} \sigma_{xb} \sigma_{xc} \cdots \sigma_{xn})$ in $V_2$ space to each basis vector. Here the $\pm$ sign produces the two different phase states in the system. Note that $e_\lambda \in S_n$; $T_\pm^{(n)} \in G_2$, and $T_\pm^{(n)}$ is a collective unitary transformation of the state.

While the basis function for each irreducible representation of $S_n$ is associated with an angular momentum state quantized along the z-direction, it does not always have the symmetry (or antisymmetry) in the spin space when the total angular momentum component $m_J \neq 0$. $T_\pm^{(n)}$ has the property to mix the $+m_J$ and $-m_J$ states and give symmetric and antisymmetric states in the $V_2$ space. It does not affect any $m_J = 0$ state as we shall see in the next Section. $T_\pm^{(n)}$ also has the physical meaning of producing a $\theta = \dfrac{\pi}{2}$ rotation of the states about the y-axis so that the new states become quantized along the x-direction.

$$\left. \begin{aligned} T_+^{(n)}|m_z\rangle &= \frac{1}{\sqrt{2}}[|m_z\rangle + |-m_z\rangle] = |m_x\rangle \\ T_-^{(n)}|m_z\rangle &= \frac{1}{\sqrt{2}}[|m_z\rangle - |-m_z\rangle] = |-m_x\rangle \end{aligned} \right\} \quad (5)$$

where $|m_i\rangle$ denotes the state quantized along the $i$th axis with component $m_i$.

The entangled pure states so produced span the $V_2^n$ space and have many interesting features useful in the identification and transformation among the states.

(2). Characterizing the tensor basis in $V_m^n$.

To characterize the constructed basis states we utilize many available criteria to test the entanglement, the separability of individual states, and the compatibility among states with $m = 2$ and different values of $n$.

(A). We use the $\rho^2 = \rho$ criterion of density matrix to test whether the state is pure.

(B). There are several ways to test the entanglement of a given state. The Peres-Horodecki separability criterion (PH)[14, 19] indicates that a state is separable from the subsystem $I$, if a matrix $\rho^{T_I}$ obtained by partial transposition of the state density matrix $\rho$ with respect to $I$ has only non-negative eigenvalues. That is, the determinant of $\rho^{T_I}$ is non-negative. This is a necessary and sufficient condition for $m = 2$, $n = 2$ case only. For other cases it is only a necessary condition.

(C). The concurrence $C_{jk} > 0$ is a necessary and sufficient condition for entanglement between a pair of qubits $j$ and $k$ [8, 10]. Concurrence is a measure of biqubit entanglement.

(D). When criterion in (B) does not give entanglement indication in the system we use another criterion which is more elaborate, but is a necessary and sufficient condition applicable to any pure state. Pope and Milburn (PM) [20] proposed a necessary and sufficient condition for the existence of genuine M-way entanglement for P-partite pure states, where P>M. It involves the test of the traces of reduced density matrix squares. $Tr\, \rho_{Q_j}^2 < 1$ for all possible sets of subsystems $Q_j$ in the M-partite subsystem if and only if the system is M-way entangled, where $\rho_{Q_j}$ is the reduced matrix obtained from the state density matrix $\rho$ by tracing over the subsystem $Q_j$. Note that our notation $\rho_{Q_j}$ is different from the one used in Ref. 13 and many other references. Under our notation, tracing over qubit $c$ in a triqubit system produces reduced matrix $\rho_c$ whereas $\rho_{ab}$ is used to denote it in Ref. 13. For multipartite systems with $n > 3$ our notation is simpler to use. We use this PM criterion for our pure states when the Peres-Horodecki criterion fails. In our examples in Sec. III it is found that this PM criterion for M-way pure state entanglement is equivalent to the criterion $\rho_{Q_j}^2 \neq \rho_{Q_j}$ in a bipartite split.

(E). When PH criterion does not show entanglement but PM shows, the system is bound entangled. The distillability of a state density matrix can be tested by examining its partial transposes. Entangled states with non-negative partial transposes cannot be distilled [12].

(F). To examine the compatibility among states with different values of $n$, we approach in two ways, comparing the angular momentum corresponding to each state and comparing the density matrices of $n$-partite states with the reduced matrices of $n+1$-partite states.

(G). The $\rho^{T_I}$ for larger $n$ is not difficult to construct if one uses the following scheme. We choose a particular set of $|N\rangle$ standard basis for the $\rho$ matrix in $n$ qubit system, here $N = 1 + \sum_{s=0}^{n} 2^s$. The standard basis are created through the numbers, 1, 2, 3,..., N, in descending order, expressed in binary form. For example, for $n = 2$, they are $|11\rangle, |10\rangle, |01\rangle, |00\rangle$; and for $n = 3$, they are $|111\rangle, |110\rangle, |101\rangle, \ldots$. Here the positions in the kets refer to qubits, $a, b, c, \ldots$, in that order. The $\rho^{T_a}$ is obtained from $\rho$ by dividing $\rho$ into 2×2 sub-matrix form and interchange the diagonal sub-matrices. The $\rho^{T_b}$ is obtained from $\rho$ by dividing $\rho$ into 4×4 sub-matrix form and interchange the diagonal sub-matrices in each quadrant. The matrix of tranposition of the $j$th qubit, $\rho^{T_j}$, is obtained from $\rho$ by dividing $\rho$ into $2^j \times 2^j$ sub-matrix form and interchange the diagonal sub-matrices in each sub-quadrant. For a subsystem $k$ including qubits $k_1, k_2, \ldots k_j$ we have the following relation $\rho^{T_k} = (((\rho^{T_{k_1}})^{T_{k_2}})\cdots)^{T_{k_j}} \equiv \rho^{T_{k_1 k_2 \ldots k_j}}$.

(H). Since the separability of a system into $M$ parts can always be examined through its separability into smaller number of parts, it is sufficient to characterize the separability of a large system through all possible division of the system into two parts [12].

(I). The degree of entanglement for each state is obtained by evaluating the concurrence and $n$-tangle.

III. Examples

For a permutation group $S_n$ the number of different permutation involving $t$ objects (*1, 2, …..t*) is $\frac{n(n-1)(n-2)\cdots(n-t+1)}{t}$, and the total number of group elements is $n!$. Therefore the number of group elements for $S_n$ increases rapidly from 2 to 24 as $n$ increases from 2 to 4. However, the construction of the multipartite wave functions is still manageable for larger number of qubits using computer algorithm.

(1). *n=2, m=2* case

There are two symmetry classes and thus two irreducible representations in $V_2^2$ space, corresponding to the first two Young's Tableaux shown in Fig. 1. In Fig. 1 the Tableaux are labeled by ($n$, $\lambda$), where $n$ denotes the number of qubits in the system and $\lambda$ the λth Tableau associated with $S_n$. The irreducible symmetrizer (primitive idempotent), $e_\lambda$, associated with each tableau $\lambda$ for $n=2$ is given, respectively, by

$$\left. \begin{array}{l} e_1 = s_2 = e + (ab) \\ e_2 = e - (ab) \end{array} \right\} \quad (6)$$

where letters are used to label the qubits, $(ab)$ denotes the permutation symmetry operation for $a$ and $b$ qubits, $s_2$ represents the group elements of $S_2$ group, and $e$ is the identity operation. Applying the $e_1$ to the biqubit basis functions, $|ab\rangle = |00\rangle, |11\rangle, |01\rangle,$ or $|10\rangle,$ where the two positions in the kets refer to qubits, $a, b$, in that order, we obtain

$$\left. \begin{array}{l} e_1 |00\rangle = |00\rangle \equiv A^+ = |2, 1, 1\rangle \\ e_1 |11\rangle = |11\rangle \equiv A^- = |2, 1, 2\rangle \\ e_1 |01\rangle = e_1 |10\rangle = \frac{1}{\sqrt{2}}(|01\rangle + |10\rangle) \equiv B^+ = |2, 1, 3\rangle \end{array} \right\} \quad (7)$$

for the first Young's tableau. We add the normalization factors for all the basis functions obtained in this work. The basis functions $(A^+, A^-, B^+)$, so obtained, are symmetric tensors belonging to the $J=1$ triplet state. They are associated with the $m_j=1, -1,$ and 0 state, respectively. We also use the notation $|n, \lambda, t\rangle$ to represent the basis function corresponding to the $t$th row within the λth irreducible representation (λth Young's Tableau) for the $n$-partitie system.

The $e_2$ operates on the $|ab\rangle$ produces a $J=0$ singlet state ($m_J=0$) for the second Young's Tableau as follows

$$\left. \begin{array}{l} e_2 |00\rangle = e_2 |11\rangle = 0 \\ e_2 |01\rangle = -e_2 |10\rangle = \frac{1}{\sqrt{2}}(|01\rangle - |10\rangle) \equiv B^- = |2, 2, 1\rangle \end{array} \right\}. \quad (8)$$

$B^-$ is an antisymmetric tensor. These basis functions, though orthogonal, do not all represent entangled states. The $A^+$ and $A^-$ states are eigenvalues of $J_z$, but are untangled.

Applying the operator $T_{\pm}^{(2)}$ further to get the symmetric and antisymmetric states in the spin space $V_2$ allows the mixing of the basis to the greatest extent. We find

$$\left.\begin{aligned} T_{\pm}^{(2)} A^+ &= T_{\pm}^{(2)} A^- = \frac{1}{\sqrt{2}}(|00\rangle \pm |11\rangle) = \Phi_{Bell}^{\pm} \\ T_{+}^{(2)} B^+ &= \frac{1}{\sqrt{2}}(|01\rangle + |10\rangle) = \Psi_{Bell}^{+} \\ T_{-}^{(2)} B^+ &= 0 \\ T_{-}^{(2)} B^- &= \frac{1}{\sqrt{2}}(|01\rangle - |10\rangle) = \Psi_{Bell}^{-} \\ T_{+}^{(2)} B^- &= 0 \end{aligned}\right\} \quad (9)$$

The $T_{\pm}^{(2)}$ operator has the physical meaning of rotating the total angular momentum quantization direction from $z$ to $x$. It is not a local unitary transformation of the form $U_a \otimes U_b$ because the maximally entangled Bell states $\Phi_{Bell}^{\pm}$ can be obtained only when all the qubits are brought together and allowed to interact. The $\Phi_{Bell}^{\pm}$ state corresponds to ($J=1$, $m_J = \pm 1$) states quantized along $x$ direction, whereas $\Psi_{Bell}^{+}$ and $\Psi_{Bell}^{-}$ correspond to ($J=1$, $m_J = 0$), and ($J=0$, $m_J=0$) state, respectively.

The Bell states are maximally entangled pure states with concurrence equal to one.

Local or global operators that may transform these entangled states among one another are summarized here. The phase bit operator, $P_x^{(2)} \equiv \sigma_{xa}\sigma_{xb}$, and the parity operator, $P_z^{(2)} \equiv \sigma_{za}\sigma_{zb}$, give the phase and parity of the Bell states, respectively, as follows.

$$\left.\begin{aligned} P_x^{(2)} \Phi_{Bell}^{\pm} &= \pm \Phi_{Bell}^{\pm} \\ P_x^{(2)} \Psi_{Bell}^{\pm} &= \pm \Psi_{Bell}^{\pm} \\ P_z^{(2)} \Phi_{Bell}^{\pm} &= \Phi_{Bell}^{\pm} \\ P_z^{(2)} \Psi_{Bell}^{\pm} &= - \Psi_{Bell}^{\pm} \end{aligned}\right\} \quad (10)$$

There also exists invertible local unitary transformations $P_{xj}^{(2)} \equiv \sigma_{xj}$ that flip the parity of the states.

$$\left.\begin{array}{l}P_{xa}^{(2)}\,\Phi_{Bell}^{\pm}=\pm\Psi_{Bell}^{\pm}\\ P_{xb}^{(2)}\,\Phi_{Bell}^{\pm}=\Psi_{Bell}^{\pm}\\ P_{xa}^{(2)}\,\Psi_{Bell}^{\pm}=\pm\Phi_{Bell}^{\pm}\\ P_{xb}^{(2)}\,\Psi_{Bell}^{\pm}=\Phi_{Bell}^{\pm}\end{array}\right\} \quad (11)$$

(2). $n=3$, $m=2$ case

The primitive idempotents for the four $n=3$ distinct Young's tableaux shown in Fig. 1 are

$$\left.\begin{array}{l}e_1 = s_3 = e + (ab) + (bc) + (ca) + (abc) + (cba)\\ e_2 = [e+(ab)][e-(ac)] = e + (ab) - (ac) - (cba)\\ e_3 = [e+(ac)][e-(ab)] = e + (ac) - (ab) - (bca)\\ e_4 = e - (ab) - (bc) - (ac) + (abc) + (cba)\end{array}\right\} \quad (12)$$

where $s_3$ represents the group elements of $S_3$ permutation group.

$e_1|abc\rangle$ produces the quartet corresponding to the total angular moment J=3/2 in the $V_2^3$ tensor space. Their basis functions are totally symmetric as given below.

$$\left.\begin{array}{l}e_1|000\rangle = |000\rangle \equiv Q_1^+ = |3,1,1\rangle\\ e_1|111\rangle = |111\rangle \equiv Q_1^- = |3,1,2\rangle\\ e_1|001\rangle = \dfrac{1}{\sqrt{3}}(|001\rangle + |010\rangle + |100\rangle) \equiv Q_2^+ = |3,1,3\rangle\\ e_1|110\rangle = \dfrac{1}{\sqrt{3}}(|110\rangle + |101\rangle + |011\rangle) \equiv Q_2^- = |3,1,4\rangle\end{array}\right\} \quad (13)$$

These functions correspond to $m_J=$3/2, -3/2, 1/2, -1/2, respectively, for $Q_1^+$, $Q_1^-$, $Q_2^+$, $Q_2^-$. Again, $|n,\lambda,\nu\rangle$ is used to denote the basis function associated with the $\lambda$th Tableau and $\nu$th basis of the $\lambda$th irreducible representation.

Each of the $e_2$ and $e_3$ produces a doublet with their basis functions symmetric with respect to qubits *a*, *b* and *a*, *c*, respectively, as shown below.

$$e_2|001\rangle = \frac{1}{\sqrt{6}}(2|001\rangle - |100\rangle - |010\rangle) \equiv D_1^+ = |3, 2, 1\rangle$$

$$e_2|110\rangle = \frac{1}{\sqrt{6}}(2|110\rangle - |011\rangle - |101\rangle) \equiv D_1^- = |3, 2, 2\rangle$$
(14)

$$e_3|001\rangle = \frac{1}{\sqrt{6}}(2|010\rangle - |100\rangle - |001\rangle) \equiv D_2^+ = |3, 3, 1\rangle$$

$$e_3|110\rangle = \frac{1}{\sqrt{6}}(2|101\rangle - |011\rangle - |110\rangle) \equiv D_2^- = |3, 3, 2\rangle$$
(15)

The total angular momentum associated with the two doublets is $J = 1/2$. $m_J$ for $D_i^+$ and $D_i^-$ are $1/2$ and $-1/2$, respectively. The partial angular momenta for these two doublets are different; $\vec{s}_a + \vec{s}_b = 1$ for the $D_1$ doublet and $\vec{s}_a + \vec{s}_c = 1$ for the $D_2$ doublet.

$e_4$ does not produce any state because there exists no totally antisymmetric tensors in $V_m^n$ if $m < n$.

Applying $T_\pm^{(3)}$ to the above generated basis functions $e_\lambda|abc\rangle$ to create symmetric and antisymmetric wave functions in the spin space, we obtain

$$T_\pm^{(3)} Q_1^\pm = \Psi_3^\pm = \frac{1}{\sqrt{2}}(|000\rangle \pm |111\rangle) = \Psi_{3GHZ}^\pm$$

$$T_\pm^{(3)} Q_2^\pm = W_3^\pm = \frac{1}{\sqrt{6}}[(|001\rangle + |010\rangle + |100\rangle) \pm (|110\rangle + |101\rangle + |011\rangle)]$$
(16)

for the quartet in $\lambda=1$ Young's Tableau;

$$T_\pm^{(3)} D_1^\pm = U_3^\pm = \frac{1}{\sqrt{12}}[(2|001\rangle - |100\rangle - |010\rangle) \pm (2|110\rangle - |011\rangle - |101\rangle)]$$
(17)

for the doublet in $\lambda=2$ Young's Tableau; and

$$T_\pm^{(3)} D_2^\pm = V_3^\pm = \frac{1}{\sqrt{12}}[(2|010\rangle - |100\rangle - |001\rangle) \pm (2|101\rangle - |011\rangle - |110\rangle)]$$
(18)

for the doublet in $\lambda=3$ Young's Tableau.

The two doublets, $D_1^\pm$, $D_2^\pm$, belong to the same Young's Diagram (but not the same Tableau) are not orthogonal, but are distinct. If we construct a doublet, which is orthogonal to $D_1^\pm$ and antisymmetric between qubit *a* and *b*, such as $F^\pm = D_1^\pm + 2D_2^\pm$, we would find very different entanglement behavior of this doublet as discussed later.

The total angular momentum and partial angular momentum designation of the three-partite states are listed in Table I. In order to study the evolution of the states from *n*-partite to (*n*+1)-partite obtained from the Young's Tableaux, we examine and compare their density matrices ρ, reduced density matrices $\rho_I$, and their classical angular momentum classifications. Firstly, we look at the *n*=2 and *n*=3 cases. We obtain unambiguous relations among these states, which can also be visualized from the simple Young's Tableaux.

By comparing ρ of the bipartite Bell states with the reduced density matrices $\rho_i$ of the 3-partite states we establish the compatibility between these two sets of states. It is listed in Table I and indicates the evolution of the states in the following way.

$$\left.\begin{array}{l} \Psi_3^\pm \leftarrow \Phi_{Bell}^\pm \\ W_3^\pm \leftarrow \Phi_{Bell}^\pm, \Psi_{Bell}^+ \\ U_3^\pm, V_3^\pm \leftarrow \Phi_{Bell}^\pm, \Psi_{Bell}^\pm \\ F^\pm \leftarrow \Psi_{Bell}^- \end{array}\right\} . \qquad (19)$$

This assignment is consistent with the angular momentum evolution and with the evolution of the Young's Tableaux when the number of qubits increases. The states generated from the λ=1 Tableau for (*n*+1)-partite are only related to those from λ=1 for smaller partite. Thus the (3, 1) is related to only the (2, 1) in Fig.1. In other words, the 3-way entangled states, $\Psi_3^\pm$, cannot be related with the singlet states, $\Psi_{Bell}^-$. On the other hand, (3, 2) or (3, 3) are related to both (2, 1) and (2, 2) in Fig.1. Since the $F^\pm$ can be expressed as the product state in biqubit *ab* and qubit *c* subspaces, $F^\pm = \Psi_{Bell}^- \otimes |\psi_c\rangle$, where $|\psi_c\rangle$ is the added qubit *c*, they can only be related to the (2, 2) bipartite Tableau in Fig. 1.

The concurrence $C_{JK}$ has been used as a useful measure of entanglement for 2-qubit and 3-qubit systems [8,11]. A residual entanglement or 3-tangle $\tau_3$ was also defined to quantify the 3-way entanglement in 3-qubit system [11]. Both concurrence and 3-tangle for general three-partite states have been examined extensively by Rajagopal and Rendell [10]. Our calculated entanglement measures for the eight states in eqs.(15-17) are listed in Table II. In Table II $C_{I(JK)}$ represents the concurrence between two subsystems; one containing the qubit *I* and the other containing biqubit *JK*. A useful measure to classify the class of states is $E_\tau \equiv C_{ab}^2 + C_{ac}^2 + C_{bc}^2$ [13]. The calculated values can also be found in Table II. For comparison with the eight states in the complete set of wave functions the corresponding results for states $F^\pm$ which do not belong to the Young's Tableau, and states $Q_2^\pm$ which do not have the symmetry in the $V_2$ space are also listed.

The separability and distillability of three-partite and multipartite systems were discussed in Ref. 14. The entanglement properties for the states given in eqs. (13-18) are examined in details in this work using various criteria mentioned in Sec.II(2). Results are listed in Table II. The states are found to be all pure states as indicated in the 2$^{nd}$ column in Table II. All states fail in the PH inseparability criteria test. Their partially transposed density matrices have no single negative eigenvalue as displayed in 5$^{th}$ column. Thus the more elaborate PM test is used to study each state. From the PM test, except the $Q_I^\pm$ states in eq.(13), all other states are found to be entangled pure states with different kinds and degrees of entanglement. Thus they belong to undistillable bound entangled states. The results for the above eight states in eqs. (15-17) are summarized in Table II. The construction of inseparable mixed states with positive partial transposition was discussed by Horodecki [24]. A new necessary and sufficient condition of separability was provided there in terms of the range of density matrices. However, such new criterion is equivalent to the old PH criterion for pure states. It is the first time that the pure states generated here are found to be inseparable but with positive partial transpositions using the PM criterion.

The first two sets of basis functions in eq.(15) are the much-discussed GHZ states and Werner states for three qubits [10]. The zero biqubit concurrence value for a given $C_{JK}$ indicates the fragilily of that state when the *I*-th qubit is removed. Here *IJK* signifies the cyclic permutation of the three qubits. The $\Psi_3^\pm$ are fragile of losing one qubit, evidenced by the zero concurrency and the zero partial transpose of its reduced density matrix. However, $\Psi_3^\pm$ is maximally three-qubit entangled with 3-tangle equal to 1 [10, 13]. Thus the operation of $T_\pm^{(3)}$ on $Q_1^\pm$ changes the states from unentangled to maximally entangled states $\Psi_3^\pm$. This GHZ state has been found for three spatially separated photons in Ref. 25, and has direct application in quantum communication and computation protocols.

$W_3^\pm$ is maximally robust against the loss of any one of the three qubits. The concurrence value for any marginal pair associated with this state is 1/3 whereas the 3-tangle is 1/3. It is interesting to see that before applying the operator $T_\pm^{(3)}$ to generate $W_3^\pm$, the state $Q_2^\pm$ has the form of $\Psi_{WRR^+}$ in Ref.10 and has concurrence 2/3 but zero 3-tangle. This is a proof that the entanglement of a linear combination of states may be different from those of individual states. Note that the $T_\pm^{(3)}$ operation increases the 3-tangle in the expense of decreasing the concurrency in $W_3^\pm$ states.

The $U_3^\pm$ ($V_3^\pm$) are symmetric with respect to qubit $a$ and $b$ ($a$ and $c$). They are also robust against the loss of any one qubit. However unlike the $W_3^\pm$ state they have different degree of entanglement among different pairs. The concurrence value for two marginal pairs is 2/3 and for one pair is 1/3. The 3-tangle is zero for these four states. The states $F^\pm$, which do not belong to any of the Young's Tableaux, are not entangled because they fail to satisfy either criterion ($\det \rho^{T_c} = 0$, $but$ $Tr \rho_c^2 = 1$). This is again a proof that the entanglement of a state may change by mixing with other states.

Any tripartite mixed state can be expressed as a linear combination of these eight states generated from the three Young's Tableaux. ($F^\pm$ states may replace the $V_3^\pm$ states in the expansion.)

From Table II we find that there is a simple way to identify the entanglement of the states without going through the much elaborate PH or PM tests. Since these states are all pure states, the fourth column can readily prove that all the states, except $F^{\pm}$, are inseparable from the subsystem $I$, because they have the relation $\rho^{T_I} \neq \rho$. $F^{\pm}$ is separable from quit $c$, but not from $a$ or $b$ since $a$ and $b$ are entangled. In addition the reduced density matrix has the unique property to identify the separability of a state because partial tracing operation gives rise the correct description of observable quantities for the state of the subsystem $I$ of multiqubit systems. In other words, it provides the correct measurement statistics for any measurement on the subsystem. For example, the 6$^{th}$ and 7$^{th}$ columns of Table II indicate that, except for $\rho_c$ of $F^{\pm}$ states, the $\rho_I^2 \neq \rho_I$ and all the values of $Tr\,\rho_I^2$ are less than 1. This shows that the subsystems are mixed states although the composite systems under consideration are pure states. The composite systems are inseparable. On the other hand, after tracing over the subsystem $c$, $F^{\pm}$ states become $\Psi_{Bell}^{-}$ state, which is a pure state. Combined with the fact that $Tr\,\rho_c^2 = 1$ it confirms the separability of the $F^{\pm}$ states into $c$ and $(ab)$ parts. The 8th column displays the fragility of $\Psi_3^{\pm}\left(F^{\pm}\right)$ after losing any one qubit (qubit $a$ or $b$).

It is interesting to analyze how the bipartite entanglement in these Tableaux changes when the number of qubits increases. The 10$^{th}$ column lists the calculated C$_{JK}$ for $n = 2, 3$. C$_{JK}$ decreases as $n$ increases. It also confirms no entanglement between $c$ and $(a, b)$ for $F^{\pm}$ although there is correlation between the two subsystems. In the 11$^{th}$ column $C_{I(JK)}$ represents the concurrence between two subsystems $I$ and $(JK)$. Our calculated result also equals to $2\sqrt{\det \rho_{JK}}$, consistent with the conjecture given in Ref. 11 for pure states. The 12$^{th}$ column displays the residual bipartite entanglement $E_\tau$ defined in Ref. 13. This value is used to show the distinct classes among the states. It is interesting to note that the states $Q_2^{\pm}$, although does not have any symmetry in the spin space, do have the maximal $E_\tau$ and averaged two-qubit concurrence. All the rest of the tripartite pure states in Table II have $E_\tau < 3/4$, in consistent with the prediction given in Ref. 13 for three-partite states. The 3-tangle for each state obtained using columns 10$^{th}$ and 11$^{th}$ are given in column 13$^{th}$. The interesting feature can be seen that the C$_{JK}$ decreases but the $\tau_3$ increases from $Q_2^{\pm}$ to $W_3^{\pm}$. There seems to be redistribution of entanglement types with the application of the $T_\pm^{(3)}$ operator. This result also shows that quantum states have finite susceptibility for entanglement [16].

The combined local or global operators $P_z^{(2)} \equiv \frac{1}{3}(\sigma_{za}\sigma_{zb} + \sigma_{zb}\sigma_{zc} + \sigma_{zc}\sigma_{za})$ and $P_x^{(3)} \equiv \sigma_{xa}\sigma_{xb}\sigma_{xc}$ serve as the phase bit and parity operators for the tripartite states.

$$\left.\begin{array}{ll} P_x^{(3)} \Psi_3^\pm = \pm \Psi_3^\pm & P_x^{(3)} \Phi_3^\pm = \pm \Phi_3^\pm \\ P_z^{(2)} \Psi_3^\pm = \Psi_3^\pm & P_z^{(2)} \Phi_3^\pm = -\Phi_3^\pm \end{array}\right\} \quad (20)$$

Here $\Phi_3^\pm$ represents $U_3^\pm$, $V_3^\pm$ or $W_3^\pm$.

These two operators have been used in the depolarization procedure to convert any mixed state into a standard form of 3-qubit density matrix with orthonomal GHZ basis in the expansion [14]. Eq.(20) gives another indication that the $\Phi_3^\pm, \Psi_3^\pm$ states cannot be depolarized and are inseparable pure states.

In addition, we find an operator $P_z^{(3)} \equiv \sigma_{za}\sigma_{zb}\sigma_{zc}$, which flips the phase and gives the parity of the states and may be called combined parity and phase flip operator.

$$\left.\begin{array}{l} P_z^{(3)} \Psi_3^\pm = \Psi_3^\mp \\ P_z^{(3)} \Phi_3^\pm = -\Phi_3^\mp \end{array}\right\} \quad (21)$$

The above operators do not change the density matrix of each state.

It is interesting to see that there are other local operators, which are non-invertible but change one state to another within the same Young's diagram. For example,

$$\left.\begin{array}{l} \frac{1}{\sqrt{3}}(\sigma_{xa} + \sigma_{xb} + \sigma_{xc})\Psi_3^\pm = W_3^\pm \\ \frac{1}{\sqrt{3}}(\sigma_{xa}\sigma_{xb} + \sigma_{xb}\sigma_{xc} + \sigma_{xc}\sigma_{xa})\Psi_3^\pm = \pm W_3^\pm \\ \sigma_{xb}\sigma_{xc} U_3^\pm = V_3^\pm \\ \sigma_{xb}\sigma_{xc} V_3^\pm = U_3^\pm \end{array}\right\} \quad (22)$$

Since one cannot convert the positivity of partial transposition of a given state by local operations [12], eqs. (11 and 22) clearly indicate that states generated from the same Young's Tableau have the same partial transposition property and same type of entanglement. Both $\Psi_3^\pm, W_3^\pm$ ($U_3^\pm, V_3^\pm$) have non-zero (zero) values of 3-entangles are the examples.

(3). *n=4, m=2* case

There are five classes in $S_4$ group characterized by their given cycle structures. Since the number of distinct Young's diagram is equal to the number of distinct irreducible representations, which is, in turn, equal to the number of classes in the group, we have five distinct diagrams for *n*=4, as shown in Fig. 1. The last two diagrams do not generate any tensor because *m<n*. Therefore we only consider the six Young's Tableaux given in Fig.1. The irreducible symmetrizer $e_\lambda$ associated with each Tableau $\lambda$ is the primitive idempotent generating the corresponding basis tensors. They are in the following forms.

$$\left.\begin{aligned}
e_1 &= s_4 = e + (ab) + (bc) + (ca) + (bd) + (cd) + (da) + (abc) + (bcd) + (dca) \\
&\quad + (dcb) + (cba) + (abd) + (acd) + (dba) + (abcd) + (dcba) + (acbd) \\
&\quad + (dbca) + (dcab) + (bacd) + (ab)(cd) + (bc)(ad) + (bd)(ac) \\
e_2 &= s_3^{(abc)}[e - (ad)] \\
e_3 &= s_3^{(abd)}[e - (ac)] \\
e_4 &= s_3^{(acd)}[e - (ab)] \\
e_5 &= [e + (ab)][e + (cd)][e - (ac)][e - (bd)] \\
e_6 &= [e + (ac)][e + (bd)][e - (ab)][e - (cd)]
\end{aligned}\right\} \quad (23)$$

where $s_4$ represents the group elements of the $S_4$ permutation group and $s_3^{(ijk)}$ represents the group elements of $S_3$ group with basis *ijk*.

Following the same procedure as in the tri-partite system, $e_\lambda |abcd\rangle$ for $\lambda$=1, 2, 3, 4, 5, 6, produces quintuplet, triplet, triplet, triplet, singlet, singlet with *J*= 2, 1, 1, 1, 0, 0, respectively. Note that each Young's Tableau is associated with a total angular momentum *J* with (2*J*+1) degenerate states.

$$\left.\begin{aligned}
&e_1|0000\rangle = |0000\rangle \equiv E^+ = |4, 1, 1\rangle \\
&e_1|1111\rangle = |1111\rangle \equiv E^- = |4, 1, 2\rangle \\
&e_1|0001\rangle = \frac{1}{2}\big[|0001\rangle + |0100\rangle + |0010\rangle + |1000\rangle\big] \equiv G^+ = |4, 1, 3\rangle \\
&e_1|1110\rangle = \frac{1}{2}\big[|1110\rangle + |1011\rangle + |1101\rangle + |0111\rangle\big] \equiv G^- = |4, 1, 4\rangle \\
&e_1|0101\rangle = \frac{1}{\sqrt{6}}\big[|0101\rangle + |1001\rangle + |0011\rangle + |0110\rangle + |1100\rangle + |1010\rangle\big] \\
&\qquad\qquad \equiv C_1^+ = |4, 1, 5\rangle
\end{aligned}\right\} \quad (24)$$

$E^+$, $E^-$, $G^+$, $G^-$, $C^+$, are states with $J=2$ and $m_J = 2, -2, 1, -1, 0$, respectively, and are symmetric with the exchange of any pair of qubits.

$$\left.\begin{aligned}
&e_2|0001\rangle = \frac{1}{2\sqrt{3}}\big[3|0001\rangle - |1000\rangle - |0010\rangle - |0100\rangle\big] \equiv L^+ = |4, 2, 1\rangle \\
&e_2|1110\rangle = \frac{1}{2\sqrt{3}}\big[3|1110\rangle - |0111\rangle - |1101\rangle - |1011\rangle\big] \equiv L^- = |4, 2, 2\rangle \\
&e_2|0101\rangle = \frac{1}{\sqrt{6}}\big[|0101\rangle + |1001\rangle + |0011\rangle - |1010\rangle - |0110\rangle - |1100\rangle\big] \\
&\qquad\qquad \equiv C_1^- = |4, 2, 3\rangle
\end{aligned}\right\} \quad (25)$$

$L^+$, $L^-$, $C_1^-$, are states with $J=1$ and $m_J = 1, -1, 0$, respectively, and are symmetric with exchange of $ab$, $bc$, $ac$ qubits.

$$\left.\begin{aligned}
&e_3|0001\rangle = \frac{1}{2\sqrt{3}}\big[3|0010\rangle - |1000\rangle - |0001\rangle - |0100\rangle\big] \equiv M^+ = |4, 3, 1\rangle \\
&e_3|1110\rangle = \frac{1}{2\sqrt{3}}\big[3|1101\rangle - |0111\rangle - |1110\rangle - |1011\rangle\big] \equiv M^- = |4, 3, 2\rangle \\
&e_3|0101\rangle = \frac{1}{\sqrt{6}}\big[|0110\rangle + |1010\rangle + |0011\rangle - |1001\rangle - |0101\rangle - |1100\rangle\big] \\
&\qquad\qquad \equiv C_2^- = |4, 3, 3\rangle
\end{aligned}\right\} \quad (26)$$

$M^+$, $M^-$, $C_2^-$, are states with $J=1$ and $m_J = 1, -1, 0$, respectively, and are symmetric with exchange of $ab$, $bd$, $ad$, qubits.

.

$$e_4|0001\rangle = \frac{1}{2\sqrt{3}}[3|0100\rangle - |1000\rangle - |0001\rangle - |0010\rangle] \equiv N^+ = |4, 4, 1\rangle$$

$$e_4|1110\rangle = \frac{1}{2\sqrt{3}}[3|1011\rangle - |0111\rangle - |1110\rangle - |1101\rangle] \equiv N^- = |4, 4, 2\rangle \quad (27)$$

$$e_4|0101\rangle = \frac{1}{\sqrt{6}}[|0110\rangle + |1100\rangle + |0101\rangle - |1001\rangle - |0011\rangle - |1010\rangle]$$

$$\equiv C_3^- = |4, 4, 3\rangle$$

$N^+$, $N^-$, $C_3^-$, are states with $J=1$ and $m_J = 1$, -1, 0, respectively, and are symmetric with exchange of *ac*, *dc*, *da* qubits.

$$e_5|0011\rangle = \frac{1}{2\sqrt{3}}\{2[|0011\rangle + |1100\rangle] - [|1001\rangle + |0110\rangle + |0101\rangle + |1010\rangle]\}$$

$$\equiv C_2^+ = |4, 5, 1\rangle \quad (28)$$

$C_2^+$, is a state with total $J=0$, $m_J = 0$, and is symmetric with simultaneous exchange of *ab*, and *dc* qubits.

$$e_6|0011\rangle = \frac{1}{2\sqrt{3}}\{2[|0101\rangle + |1010\rangle] - [|1001\rangle + |0110\rangle + |0011\rangle + |1100\rangle]\}$$

$$\equiv C_3^+ = |4, 6, 1\rangle \quad (29)$$

$C_3^+$, is a state with total $J=0$, $m_J = 0$, and is symmetric with simultaneous exchange of *ac*, and *bd* qubits.

Applying $T_\pm^{(4)}$ to the generated basis $e_\lambda|abcd\rangle$ we obtain

$$T_\pm^{(4)} E^\pm = \Psi_4^\pm = \frac{1}{\sqrt{2}}(|0000\rangle \pm |1111\rangle) = \frac{1}{\sqrt{2}}(E^+ \pm E^-) = \Psi_{4GHZ}^\pm$$

$$T_\pm^{(4)} G^\pm = W_4^\pm = \frac{1}{\sqrt{2}}[G^+ \pm G^-] \quad (30)$$

$$T_\pm^{(4)} C_1^+ = C_1^+ = \frac{1}{\sqrt{6}}[|0101\rangle + |1001\rangle + |0011\rangle + |0110\rangle + |1100\rangle + |1010\rangle]$$

for the $\lambda=1$ Tableau;

$$T_\pm^{(4)} L^\pm = X_4^\pm = \frac{1}{\sqrt{2}}[L^+ \pm L^-]$$

$$\quad (31)$$

$$T_\pm^{(4)} C_1^- = C_1^- = \frac{1}{\sqrt{6}}[|0101\rangle + |1001\rangle + |0011\rangle - |1010\rangle - |0110\rangle - |1100\rangle]$$

for the λ=2 Tableau;

$$T_\pm^{(4)} M^\pm = Y_4^\pm = \frac{1}{\sqrt{2}}[M^+ \pm M^-]$$

$$T_\pm^{(4)} C_2^- = C_2^- = \frac{1}{\sqrt{6}}[|0110\rangle + |1010\rangle + |0011\rangle - |1001\rangle - |0101\rangle - |1100\rangle]$$

(32)

for the λ=3 Tableau;

$$T_\pm^{(4)} N^\pm = Z_4^\pm = \frac{1}{\sqrt{2}}[N^+ \pm N^-]$$

$$T_\pm^{(4)} C_3^- = C_3^- = \frac{1}{\sqrt{6}}[|0110\rangle + |1100\rangle + |0101\rangle - |1001\rangle - |0011\rangle - |1010\rangle]$$

(33)

for λ=4 Tableau;

$$T_\pm^{(4)} C_2^+ = C_2^+ = \frac{1}{2\sqrt{3}}\{2[|0011\rangle + |1100\rangle] - [|1001\rangle + |0110\rangle + |0101\rangle + |1010\rangle]\}$$ (34)

for λ=5 Tableau; and

$$T_\pm^{(4)} C_3^+ = C_3^+ = \frac{1}{2\sqrt{3}}\{2[|0101\rangle + |1010\rangle] - [|1001\rangle + |0110\rangle + |0011\rangle + |1100\rangle]\}$$ (35)

for λ=6 Tableau.

The application of $T_\pm^{(4)}$ does not affect the $C_j^\pm$ states which are $m_J = 0$ states. This is consistent with the physical meaning of the $T_\pm^{(n)}$ being the rotation operator of the quantization axis for the total angular momentum.

These sixteen wave functions in eqs.(30-35) form the complete basis for four-qubit system. Any state in this system may be expanded in terms of these wave functions. They are entangled pure states because each one of them cannot be expressed as a direct product state. The total angular momentum and partial angular momentum designations of these states are listed in Table III.

The relationship between the $n = 4$ and $n = 3$ states are given as follows.

$$\left.\begin{array}{l}\Psi_4^\pm \leftarrow \Psi_3^\pm \\ W_4^\pm, C_1^\pm, X_4^\pm \leftarrow \Psi_3^\pm, W_3^\pm \\ Y_4^\pm, C_2^\pm \leftarrow U_3^\pm \\ Z_4^\pm, C_3^\pm \leftarrow V_3^\pm\end{array}\right\} \quad (36)$$

These relations are again established by comparing the $\rho$ of the 3-partite states with the reduced density matrices $\rho_I$ of the 4-partite states, and by considering the angular momentum evolution of respective states. This result is consistent with the simple intuitive argument from the structure of the corresponding Young's Tableau.

We have also adapted our method in Sec. II(2) to test the entanglement of the states in the 4-qubit system here. All the basis state density matrices do not have negative transposition under the PH test. But they all pass the test of PM to indicate that they are bound entangled states. They are also undistillable. The classification of the separability for general 4-qubit systems has been discussed in Ref. 12. Although there are over 300 different ways that this system can be separated, only one level of the hierarchic structure, namely the 2-separability, is required to completely classify the states in this system. This is based on the fact that $n$-separability of the system split follows from the corresponding ($n$-1)-separability. Thus we only need to study the 2-separability of all possible two-part splits in Table IV in order to determine the entanglement of each state.

In Table IV, $I$ represents one of the split-off subsystem which contains a single qubit or biqubit, and $J$ represents any single qubit that is not in the subsystem $I$. The first column lists the states from different Young's Diagrams. Results for states from different Young's Tableaux can be obtained from their corresponding states in the same Diagram through the exchange of labels given in Fig. 1. Thus results for $Y_4^\pm$, $Z_4^\pm$ can be obtained from that for $X_4^\pm$; $C_2^-$, $C_3^-$ from that for $C_1^-$; and $C_3^+$ from that for $C_2^+$ in Table IV. The result in second column indicates that all the states in the Table are pure states. The third column list some possible subsystems $I$ that characterize different entanglement behavior for each state, a complete list of the subsystems can be obtained through the symmetry of qubits given in Table III. The fourth column shows $\rho^{T_I} \neq \rho$ for all the states. This criterion alone is sufficient to show the entanglement of these pure states. However, in order to double check with other criteria using PH and PM tests we calculate the transpositions of the density matrices and their reduced density matrices. Results are also listed in Table IV. The fifth column clearly shows that PH test is not sufficient to prove the entanglement of these pure states. The sixth column indicates that the reduced matrices do not come from pure states. The seventh column gives the successful test of entanglement by PM test, which is $Tr\ \rho_I^2 < 1$ for all $I$. This criterion is in fact equivalent to the separability criterion given by Wang [26]. He defined a polarized vector $\vec{\xi}_I$ from the reduced density matrix $\rho_I$. The condition $|\vec{\xi}_I|^2 = 1$ leads to $Tr\ \rho_I^2 = 1$. The eighth column shows that, besides the first two states, all the states are robust after removing the subsystem $I$. The first two states, similar to the case of $\Psi_3^\pm$ in Table II, have $\rho_I^{T_J} = \rho_I$ displaying the separability of the remaining system after $I$ is removed. An unambiguous conclusion can therefore be drawn from all the different tests.

The $\Psi_4^\pm$ states are the GHZ states and are found to be 4-way entangled. They are most fragile toward the removing of any one qubit. The $W_4^\pm$, unlike the $W_3^\pm$ counter part, are not the maximally robust states against the removing of any qubit because $\rho_I^{T_J} = \rho_I$. This is consistent with the fact that this state is related to the fragile state $\Psi_3^\pm$ as shown in Table III. The rest of the states in eqs.(30-35) are robust against the removing of any single qubit.

Four-partite states have been discussed by many authors [15, 16, 27 ]. Smolin constructed 4-qubit state through two pairs of entangled 2-quit states [15]. Lee *et al.* also constructed 4-qubit states from two pairs of 2-qubit states based on the angular momentum addition formalism [27]. Some of the states constructed are separable states. Verstraete *et al.* considered nine families of states and used the concept of concurrence and mixed 3-tangle to analyze those states [16]. Unlike the present construction the symmetries in the particle space and spin space were not simultaneously taken into account in the previous works.

The potential general entanglement measure has been proposed by Wong and Christensen [28]. They define "$n$-tangle $\tau_n$" for even number of $n$ greater than 3.

$$\left. \begin{array}{l} \tau_n \equiv C^2_{1,2....n} = \left|\langle\psi|\tilde{\psi}\rangle\right|^4 \\ |\tilde{\psi}\rangle = \sigma_y^{\otimes n}|\psi^*\rangle \end{array} \right\} \tag{37}$$

where $|\psi^*\rangle$ is the complex conjugate of the $n$ qubit state $|\psi\rangle$, and $\sigma_y^{\otimes n}$ denotes the spin flip operator in the $n$-qubit system. For $n = 2$, the expression in eq.(37) coincides with that of concurrence.

The calculated $\tau_4$ for our 4-qubit systems are given in the 4$^{th}$ column in Table V. Like the biqubit case, all the "4-tangles" equal to one for states with symmetry in spin space, and equal to zero for states without the symmetry. $A^\pm$ in eq.(7), $E^\pm$, $G^\pm$, $L^\pm$, $M^\pm$, and $N^\pm$ in eqs.(24-27) belong to the latter. This further demonstrates the important correlation between the "$n$-tangle" and the symmetry in spin space for even $n$. One may conjecture that only the states generated using the present method giving the required symmetry in spin space would all have maximum "$n$-tangle" for even $n$. The odd $n$ systems have peculiar behavior. Their $n$-tangles have not yet been defined when $n > 3$. However, even 3-tangle shows its increase in value when the symmetry in spin space is present as discussed in Sec. III(2). The "$n$-tangle" so defined is not by itself a measure of $n$-way entanglement as already mentioned by Wong and Christensen [28]. We find that the state $R \equiv C_2^+ - C_3^+ = |0011\rangle + |1100\rangle - |1001\rangle - |0110\rangle$, which is a two-subsystem ($ac$)-($bd$) separable state, has "4-tangles" equal to one also. By examining the forms of $R$ and eq.(37), it becomes obvious that $\tau_n$ defined in eq.(37) can accurately measure the degree of spin space symmetry in a state for even $n$, but not the $n$-way entanglement. How to define a $n$-tangle that can uniquely measure the $n$-way entanglement for $n > 3$ within multiqubit systems remains to be explored in the near future.

There has not been any other alternative method to quantify the degree of entanglement for $n > 3$. The relative degree of entanglement for the states obtained here for $n = 4$ is further analyzed in detail by examining the concurrence of two-subsystem split. The reduced matrices $\rho_I \neq \rho_I^2$ in the 6$^{th}$ column of Table IV indicate that the corresponding state after tracing over one subsystem is no longer a pure state. Thus there is no direct way to evaluate the $C_{I(JK)}$ and $\tau_3$. Listed in Table V are the evaluated values for $C_{JK}$ and $C_{I(JKL)}$ only. $C_{JK}$ represents the concurrence between any pair of qubits in the system unless specified by the particular pairs inside parenthesis. $C_{I(JKL)}$ represents the concurrence between the remaining (*JKL*) subsystem and any qubit, *I*, unless specified by the particular single qubit inside the parenthesis. $C_{I(JKL)}$ is equal to $2\sqrt{\det \rho_{JKL}}$ for the states under consideration. In order to compare the states with and without the symmetry in the spin space $V_2$, we also evaluate the concurrences for the $E^\pm, G^\pm, L^\pm$ and $R$ states. As shown in Table V, there exists redistribution of entanglement when the symmetry sets in. In particular, the $C_{I(JKL)}$ increases with the symmetry and reaches maximum value when the pure states $E^\pm, G^\pm, L^\pm$ change into $\Psi_4^\pm, W_4^\pm, X_4^\pm$, respectively. The state *R*, like the *F* in *n* = 3, is separable, but has maximum concurrence for $C_{I(JKL)}$ and for some of the $C_{JK}$.

Based on the above analysis one may suggest that the symmetry in both the particle and spin spaces plays a significant role in the distribution and degree of entanglement in multipartite systems.

The local operators $P_x^{(4)} \equiv \sigma_{xa}\sigma_{xb}\sigma_{xc}\sigma_{xd}$ and $P_z^{(4)} \equiv \sigma_{za}\sigma_{zb}\sigma_{zc}\sigma_{zd}$ also serve as the phase bit and parity operators for the 4-partite states.

$$\left.\begin{array}{l} P_x^{(4)} \Psi_4^\pm = \pm \Psi_4^\pm \quad P_x^{(4)} \Phi_4^\pm = \pm \Phi_4^\pm \quad P_x^{(4)} C_i^\pm = \pm C_i^\pm \\ P_z^{(4)} \Psi_4^\pm = +\Psi_4^\pm \quad P_z^{(4)} \Phi_4^\pm = -\Phi_4^\pm \quad P_z^{(4)} C_i^\pm = +C_i^\pm \\ i = 1, 2, 3; \qquad \Phi_4 \text{ are } W_4, X_4, Y_4, Z_4. \end{array}\right\} \quad (38)$$

However, relations similar to eqs. (11 and 22) cannot be found through local operator transformations for *n* = 4.

Experimental demonstrations of four-partite entanglements were reported for photons [29-31] and for ions [32]. Theoretical construction of bound entangled states for multipartite systems is therefore helpful to the future design of devices in quantum computing and teleportation.

IV. Summary

We propose to use Young's Tableaux to generate complete sets of basis states for multi-partite systems. These basis states possess the permutation symmetry in particle space $V^n$ given by the idempotent associated with each Tableau. They can also possess the required symmetry in the $m$-dimensional space associated with each partite by operating on with another appropriate operator in the $V_m$ space, such as the $T_\pm^{(n)}$ operator in spin space $V_2$. In particular we apply this method to examine the $m = 2$, $n = 2, 3, 4$ qubit systems. By analyzing the characteristics of each state we obtain the following results.

(1). The states generated from a given Young's Tableau have the same total angular momentum but with different component $m_z$. The states generated from each Young's Tableau originated from the same Young's Diagram have different partial angular momenta, and they are distinct but non-orthogonal states. States obtained from all the Young's Tableaux for a given $n$ form a complete set of wave functions for the $V_m^n$ space.

(2). The $T_\pm^{(n)}$ operator we defined in the spin space of the $n$-qubit systems changes the unentangled states $A^\pm$, $Q^\pm$, $E^\pm$, in $n = 2, 3, 4$ cases, into maximally entangled states $\Phi_{Bell}^\pm$, $\Psi_{3GHZ}^\pm$, $\Psi_{4GHZ}^\pm$, respectively. It also increases the degree of n-way entanglement of many other states. The $T_\pm^{(n)}$ operator has the physical meaning of rotating the quantizing axis for the total angular momentum from $z$ to $x$-axis and gives rise symmetry in spin space.

(3). The states generated from the $\lambda=1$ Tableau for $(n+1)$- partite are only related to those from the $\lambda=1$ Tableau for smaller partites. From the compatibility relation among other states listed in Tables I and III, the evolution of these states is consistent with simple graphical relations among the Young's Tableaux. For example, in Fig. 1, Tableau (4, 1) is related to (3, 1) and (2, 1), Tableaux (4, 5) and (4, 3) are related with (3, 2); Tableaux (4, 4) and (4, 6) are related with (3, 3); etc.

(4). All the states generated are undistillable bound entangled states. This follows from the criterion that a maxially entangled state $\rho_N$ can be distilled if and only if all possible bipartite splits of the N qubits have negative partial transposition [12]. The density matrices for $n > 2$ cases do not have negative partial transpositions and thus cannot be tested using the Pere-Horodecki criterion [19, 14]. However, more elaborate procedures given by Pope-Milburn criterion [20] are used to test the inseparability of those states. From the tested results we conclude that for the states we generate the criterion $\rho^{T_I} \neq \rho$ is a sufficient condition for state entanglement. We are willing to conjecture that all the states generated for $n > 4$ qubit systems following our procedure are also undistillable bound entangled states.

(5). States generated from the same Young's Tableau have the same partial transposition property.

(6). The separability of our $n$-partite pure states are completely determined by their two-partite reduced density matrices in Tables II and IV. This is consistent with the conjecture given in Ref. 33.

(7). The calculated "$n$-tangles" for our $n = 2$ and 4 states are all equal to one signifying the complete spin space symmetry. Our conjecture is that this will hold for all states generated from Young's Tableaux for even $n$. However, "$n$-tangles" does not measure $n$-way entanglement. The relative degree of entanglement for the states obtained here for $n = 4$ remains to be determined in the future.

(8). Symmetry plays a significant role in the distribution of entanglement among the qubits.

ACKNOWLEDGEMENTS
The author would like to thank Dr. A. K. Rajagopal for useful discussions. This work is supported in part by the Office of Naval Research.

Table I

Angular momentum associated with each 2-qubit and 3-qubit state.
($\vec{s}_i$ represents the spin state of the $i$-th qubit, and J represents the total angular momentum)

| state | $s_i$ | J | $m_J$ | Compatibility | Symmetric w.r.t. |
|---|---|---|---|---|---|
| $\Phi_{Bell}^{\pm}$ | | 1 | $\pm 1$ | | $a, b$ |
| $\Psi_{Bell}^{+}$ | | 1 | 0 | | $a, b$ |
| $\Psi_{Bell}^{-}$ | | 0 | 0 | | $a, b$ |
| $\Psi_3^{\pm}$ | $\vec{s}_a + \vec{s}_b = 1$ | $\frac{3}{2}$ | $\pm\frac{3}{2}$ | $\Phi_{Bell}^{\pm}$ | $a, b, c$ |
| $W_3^{\pm}$ | $\vec{s}_a + \vec{s}_b = 1$ | $\frac{3}{2}$ | $\pm\frac{1}{2}$ | $\Phi_{Bell}^{\pm}, \Psi_{Bell}^{+}$ | $a, b, c$ |
| $U_3^{\pm}$ | $\vec{s}_a + \vec{s}_b = 1$ | $\frac{1}{2}$ | $\pm\frac{1}{2}$ | $\Phi_{Bell}^{\pm}, \Psi_{Bell}^{\pm}$ | $a, b$ |
| $V_3^{\pm}$ | $\vec{s}_a + \vec{s}_c = 1$ | $\frac{1}{2}$ | $\pm\frac{1}{2}$ | $\Phi_{Bell}^{\pm}, \Psi_{Bell}^{\pm}$ | $a, c$ |
| $F^{\pm}$ | $\vec{s}_c + \vec{s}_b = 0$ | $\frac{1}{2}$ | $\pm\frac{1}{2}$ | $\Psi_{Bell}^{-}$ | $b, c$ |

Table II
Entanglement tests with each 2-qubit and 3-qubit state.

($\rho^{T_I}$ represents the partial transposition of the density matrix $\rho$ with respect to the subsystem $I$ which is one of the qubits list in the third column, and $\rho_I$ represents the reduced matrix tracing over the subsystem $I$ of $\rho$. $C_{JK}$ is the concurrence of two-qubit entanglement. $IJK$ represents the cyclic permutation of the qubits. $\tau_3$ represents 3-tangle.)

| state | $\rho^2$ | $I$ | $\rho^{T_I}$ | $\det \rho^{T_I}$ | $\rho_I^2$ | $\mathrm{Tr}\,\rho_I^2$ | $\rho_I^{T_J}$ | $\det \rho_I^{T_J}$ | $C_{JK}$ | $C_{I(JK)}$ | $E_\tau$ | $\tau_3$ | |
|---|---|---|---|---|---|---|---|---|---|---|---|---|---|
| $\Phi_{Bell}^\pm$ | $\rho$ | $a, b$ | $\neq \rho$ | $-\frac{1}{2}$ | $\neq \rho_I$ | $\frac{1}{2}$ | | | 1 | | | | maximally entangled |
| $\Psi_{Bell}^\pm$ | $\rho$ | $a, b$ | $\neq \rho$ | $-\frac{1}{2}$ | $\neq \rho_I$ | $\frac{1}{2}$ | | | 1 | | | | maximally entangled |
| $\Psi_3^\pm$ | $\rho$ | $a, b, c$ | $\neq \rho$ | 0 | $\neq \rho_I$ | $\frac{1}{2}$ | $= \rho_I$ | 0 | 0 | 1 | 0 | 1 | bound entangled, fragile |
| $W_3^\pm$ | $\rho$ | $a, b, c$ | $\neq \rho$ | 0 | $\neq \rho_I$ | $\frac{13}{18}$ | $\neq \rho_I$ | <0 | $\frac{1}{3}$ | $\frac{\sqrt{5}}{3}$ | $\frac{1}{3}$ | $\frac{1}{3}$ | bound entangled, robust |
| $U_3^\pm$ | $\rho$ | $c$ | $\neq \rho$ | 0 | $\neq \rho_I$ | $\frac{5}{9}$ | $\neq \rho_I$ | <0 | $\frac{1}{3}$ | $\frac{2\sqrt{2}}{3}$ | 1 | 0 | (bound entangled, symmetric |
| | | $a, b$ | $\neq \rho$ | 0 | $\neq \rho_I$ | $\frac{13}{18}$ | $\neq \rho_I$ | <0 | $\frac{2}{3}$ | $\frac{\sqrt{5}}{3}$ | | | with respect to qubit $a$ and $b$) |
| $V_3^\pm$ | $\rho$ | $b$ | $\neq \rho$ | 0 | $\neq \rho_I$ | $\frac{5}{9}$ | $\neq \rho_I$ | <0 | $\frac{1}{3}$ | $\frac{2\sqrt{2}}{3}$ | 1 | 0 | (bound entangled, symmetric |
| | | $a, c$ | $\neq \rho$ | 0 | $\neq \rho_I$ | $\frac{13}{18}$ | $\neq \rho_I$ | <0 | $\frac{2}{3}$ | $\frac{\sqrt{5}}{3}$ | | | with respect to $a$ and $c$) |
| $F^\pm$ | $\rho$ | $c$ | $= \rho$ | 0 | $= \rho_I$ | 1 | $\neq \rho_I$ | <0 | 1 | 0 | 1 | 0 | (unentangled, and |
| | | $a, b$ | $\neq \rho$ | 0 | $\neq \rho_I$ | $\frac{1}{2}$ | $= \rho_I$ | <0 | 0 | 1 | | | antisymmetric $w.r.t.$ $a$ and $b$) |
| $Q_2^\pm$ | $\rho$ | $a, b, c$ | $\neq \rho$ | 0 | $\neq \rho_I$ | $\frac{5}{9}$ | $\neq \rho_I$ | <0 | $\frac{2}{3}$ | $\frac{2\sqrt{2}}{3}$ | $\frac{4}{3}$ | 0 | bound entangled, robust |

Table III

Angular momentum associated with each 4-qubit state.
($s_i$ represents the spin state of the $i$-th qubit, and J represents the total angular momentum)

| state | $s_i$ | J | $m_J$ | Compatibility | Symmetric w.r.t. |
|---|---|---|---|---|---|
| $\Psi_4^\pm$ | $\vec{s}_a + \vec{s}_b + \vec{s}_c = \frac{3}{2}$ | 2 | $\pm 2$ | $\Psi_3^\pm$ | a, b, c, d |
| $W_4^\pm$ | $\vec{s}_a + \vec{s}_b + \vec{s}_c = \frac{3}{2}$ | 2 | $\pm 1$ | $\Psi_3^\pm, W_3^\pm$ | a, b, c, d |
| $C_1^+$ | $\vec{s}_a + \vec{s}_b + \vec{s}_c = \frac{3}{2}$ | 2 | 0 | $W_3^\pm$ | a, b, c, d |
| $X_4^\pm$ | $\vec{s}_a + \vec{s}_b + \vec{s}_c = \frac{3}{2}$ | 1 | $\pm 1$ | $\Psi_3^\pm, W_3^\pm$ | a, b, c |
| $C_1^-$ | $\vec{s}_a + \vec{s}_b + \vec{s}_c = \frac{3}{2}$ | 1 | 0 | $W_3^\pm, U_3^\pm$ | a, b, c |
| $Y_4^\pm$ | $\vec{s}_a + \vec{s}_b + \vec{s}_d = \frac{1}{2}$ | 1 | $\pm 1$ | $U_3^\pm$ | a, b, d |
| $C_2^-$ | $\vec{s}_a + \vec{s}_b + \vec{s}_d = \frac{1}{2}$ | 1 | 0 | $U_3^\pm$ | a, b, d |
| $Z_4^\pm$ | $\vec{s}_a + \vec{s}_c + \vec{s}_d = \frac{1}{2}$ | 1 | $\pm 1$ | $V_3^\pm$ | a, c, d |
| $C_3^-$ | $\vec{s}_a + \vec{s}_c + \vec{s}_d = \frac{1}{2}$ | 1 | 0 | $V_3^\pm$ | a, c, d |
| $C_2^+$ | $\vec{s}_a + \vec{s}_b = \vec{s}_c + \vec{s}_d = 1$ | 0 | 0 | $U_3^\pm$ | (a, b), (c, d) |
| $C_3^+$ | $\vec{s}_a + \vec{s}_c = \vec{s}_b + \vec{s}_d = 1$ | 0 | 0 | $V_3^\pm$ | (a, c), (b, d) |

Table IV

Entanglement tests with each 4-qubit state.

($\rho^{T_I}$ represents the partial transposition of the density matrix $\rho$ with respect to the subsystem $I$ which is one of the subsystems list in the third column, and $\rho_I$ represents the reduced matrix tracing over the subsystem $I$ of $\rho$. $J$ is a single qubit that does not belong in the subsystem $I$.)

| state | $\rho^2$ | $I$ | $\rho^{T_I}$ | det $\rho^{T_I}$ | $\rho_I^2$ | Tr $\rho_I^2$ | $\rho_I^{T_J}$ | det $\rho_I^{T_J}$ |
|---|---|---|---|---|---|---|---|---|
| $\Psi_4^\pm$ | $\rho$ | $a, b, c, d, ab, ac, ad$ | $\neq \rho$ | 0 | $\neq \rho_I$ | $\frac{1}{2}$ | $= \rho_I$ | 0 |
| $W_4^\pm$ | $\rho$ | $a, b, c, d, ab, ac, ad$ | $\neq \rho$ | 0 | $\neq \rho_I$ | $\frac{1}{2}$ | $= \rho_I$ | 0 |
| $C_1^+$ | $\rho$ | $a, b, c, d$ | $\neq \rho$ | 0 | $\neq \rho_I$ | $\frac{1}{2}$ | $\neq \rho_I$ | 0 |
|  |  | $ab, ac, ad$ | $\neq \rho$ | 0 | $\neq \rho_I$ | $\frac{1}{2}$ | $\neq \rho_I$ | $< 0$ |
| $X_4^\pm$ | $\rho$ | $a, b, c, d$ | $\neq \rho$ | 0 | $\neq \rho_I$ | $\frac{1}{2}$ | $\neq \rho_I$ | 0 |
|  |  | $ad, bd, cd, ba, bc$ | $\neq \rho$ | 0 | $\neq \rho_I$ | $\frac{1}{2}$ | $\neq \rho_I$ | $< 0$ |
| $C_1^-$ | $\rho$ | $a, b, c, d$ | $\neq \rho$ | 0 | $\neq \rho_I$ | $\frac{1}{2}$ | $\neq \rho_I$ | 0 |
|  |  | $cd, ba$ | $\neq \rho$ | 0 | $\neq \rho_I$ | $\frac{1}{2}$ | $\neq \rho_I$ | $< 0$ |
| $C_2^+$ | $\rho$ | $cd, ab$ | $\neq \rho$ | 0 | $\neq \rho_I$ | $\frac{1}{3}$ | $\neq \rho_I$ | $> 0$ |
|  |  | $c, d$ | $\neq \rho$ | 0 | $\neq \rho_I$ | $\frac{1}{2}$ | $\neq \rho_I$ | $> 0$ |
|  |  | $a, b$ | $\neq \rho$ | 0 | $\neq \rho_I$ | $\frac{1}{2}$ | $\neq \rho_I$ | $= 0$ |
|  |  | $ad, bd, ac, bc$ | $\neq \rho$ | 0 | $\neq \rho_I$ | $\frac{7}{12}$ | $\neq \rho_I$ | $< 0$ |

Table V

Concurrences of two-partites in each 4-qubit state.

($C_{JK}$ represents the concurrence between any pair of qubits in the system unless specified by the parenthesis $(JK)$. $C_{I(JKL)}$ represents the concurrence between any qubit $I$ and the rest of the system $JKL$ unless specified by $(I)$. $\tau_4$ represents 4-tangle.)

| state | $C_{JK}$ | $C_{I(JKL)}$ | $\tau_4$ |
|---|---|---|---|
| $\Psi_4^\pm$ | 0 | 1 | 1 |
| $W_4^\pm$ | 0 | 1 | 1 |
| $C_1^+$ | $\frac{1}{3}$ | 1 | 1 |
| $X_4^\pm$ | $\frac{1}{3}$ | 1 | 1 |
| $C_1^-$ | $\frac{1}{3}$ | 1 | 1 |
| $C_2^+$ | 0 $(cd, ab)$, $\frac{1}{2}$ $(ac, bc, ad, bd)$ | 1 | 1 |
| $E^\pm$ | 0 | 0 | 0 |
| $G^\pm$ | $\frac{1}{2}$ | $\frac{\sqrt{3}}{2}$ | 0 |
| $L^\pm$ | $\frac{1}{2}$ $(ad, bd, cd)$, $\frac{1}{6}$ $(ab, bc, ac)$ | $\frac{\sqrt{3}}{2}$ $(d)$, $\frac{\sqrt{11}}{6}$ $(a, b, c)$ | 0 |
| $R$ | 0 $(ad, bc, ab, cd)$, 1 $(ac, bd)$ | 1 | 1 |

Figure caption

Fig.1. Young's Tableaux for permutation groups $n = 2$, 3, and 4. The label for each Tableau is designated by $(n, \lambda)$.

| a | b |

(2, 1)

| a |
| b |

(2, 2)

$(n, \lambda)$

| a | b | c |

(3, 1)

| a | b |
| c |

(3, 2)

| a | c |
| b |

(3, 3)

| a |
| b |
| c |

(3, 4)

| a | b | c | d |

(4, 1)

| a | b | c |
| d |

(4, 2)

| a | b | d |
| c |

(4, 3)

| a | c | d |
| b |

(4, 4)

| a | b |
| c | d |

(4, 5)

| a | c |
| b | d |

(4, 6)

| a | d |
| b |
| c |

| a |
| b |
| c |
| d |

31